\documentstyle [12pt] {article}
\title{ZENO EFFECT (FREQUENT INHIBITION OF THE EFFECTOR CELLS REGULATORS) IN CANCER THERAPY}

\author{Rade Glavatovi\'c$^{\sharp}$,Vladan Pankovi\'c$^{\ast}$\\
$^\sharp$ Military-Medical Academy, 11000 Belgrade, Crnotravska
17, Serbia \\
$^\ast$Department of Physics, Faculty of Sciences,
21000 Novi
Sad,\\
Trg Dositeja Obradovi\'ca 4. , Serbia, vpankovic@if.ns.ac.yu \\
 \\}

\date {}
\begin{document}
\maketitle


\begin {abstract}
Recently Brutovsky and Horvath suggested a strategy of pure
evolutionary self-destroying of the cancer without any active
medical treatment. In this work we suggest a completely opposite
strategy for cancer inhibition and eventually elimination. It is
based by frequent (many times repeated) application of an especial
active medical treatment. This treatment represents such
inhibition of the regulator cells ($Th_{1}$, $Th_{2}$, …) which
cause hyper-activity of the effector cells (citotoxic limfocits,
nature killer cells,  …) that eliminate cancer cells (dirty
inspector Harry effect). Conceptually, our strategy is similar to
Zeno effect theoretically predicted and experimentally verified in
the quantum mechanics (but which can be realized in practically
any domain of the physics). According to Zeno effect a non-stable
system, that during time evolves from initial non-decayed in the
final decayed state can never decay by frequent (many times
repeated) perturbation by measurement (representing an active
evolution breaking treatment).
\end {abstract}

{\it "Neka bude \v{s}to biti ne mo\v{z}e!}\\
\makebox[5cm][l]{neka bude borba neprestana -"}

\vspace{0.3cm}

{"let it be what cannot be!\\
Let it be the neverending struggle-"}

\vspace{0.3cm}

Petar II Petrovi\'c Njego\v{s}, "The Mountain Wreath"\\

\vspace {0.8cm}

"Sie r\"{u}ckt und weicht, der Tag is \"{u}berlebt,\\
 Dort eilt sie hin und földert neues Leben.\\
 Oh, das kein Flügel mich vom Boden hebt,\\
 Ihr nach und immer nach zu streben!"\\

 \vspace{0.3cm}

("The glow retreats, done is the day of toil;\\
  It yonder hastes, new fields of life exploring;\\
  Ah, that no wing can lift me form the soil,\\
  Upon its track to follow, follow soaring!")\\

              Jochan Wolfgang G\"{o}te, "Faust"

\vspace {1.5cm}

\section {Introduction}
Brutovsky and Horvath [1] suggested recently an original strategy
against cancer disease. They started, on the one hand, from
unambiguous fact that active medical treatment (chirurgic
treatment, radiation or chemo-therapy) in a late phase of the
disease can often cause, instead of disease inhibition, late
accelerated disease expansion. In this sense active medical
treatment behaves like a penalty function. On the other hand
Brutovsky and Horvath supposed that cancer growth can be
considered as a dynamical evolution of the cancer cells
population. This evolution, as an optimization strategy, becomes
during time les and les efficient (or, cancer cells affect the
normal cells les and les efficiently) which can finally cause a
complete elimination of the disease. More precisely, according to
some relevant theoretical analyses [2], [3] on the carcinogenesis
Brutovsky and Horvath suggested an abstract theoretical mechanism
"keeping in mind an eventual therapeutic application" and "focus
on those aspect of evolutionary optimization which decrease or
inhibit efficiency of the optimization process", even if "strict
adherence to the optimization framework has lead us to
counterintuitive implications" on the active medical treatment as
penalty function. Thus, Brutovsky and Horvath suggested a strategy
of pure evolutionary self-destroying of the cancer without any
active medical treatment.

In this work we shall suggest a completely opposite strategy for
cancer inhibition and eventually elimination. It is based by
frequent (many times repeated) application of an especial active
medical treatment. This treatment represents such inhibition of
the regulator cells ($Th_{1}$, $Th_{2}$, …)  which cause
super-activity of the effector cells (citotoxic limfocits, nature
killer cells, …) that eliminate cancer cells (dirty inspector
Harry effect). All this represents further development of our
previous ideas and observations [4] on the opposite functioning of
the cancer and hyper-immune diseases, e.g. multiple sclerosis.
Conceptually, our strategy is similar to Zeno effect theoretically
predicted [5] and experimentally verified [6], [7] in the quantum
mechanics, but which can be realized in practically any domain of
the physics [8]. According to Zeno effect a non-stable system,
that during time dynamically evolves from initial non-decayed in
the final decayed state, can never decay by frequent (many times
repeated) perturbation of the dynamics by measurement
(representing an active dynamical evolution breaking treatment).

\section {Zeno effect (frequent inhibition of the effector cells regulators) in cancer therapy}

Brutovsky and Horvath strategy against cancer can be simplifiedly
presented by the following population dynamics equation
\begin {equation}
    \frac {dp}{dt} = ({\it a}-{\it b}t)p
\end {equation}
with simple solution
\begin {equation}
    p = p_{0} \exp[{\it a}t- \frac {1}{2}{\it b}t^{2}]                     .
\end {equation}
Here $p$ represents the cancer cells population in the time moment
$t$, $p_{0}$ - initial cancer cells population, ${\it a}$ - time
independent cancer cells population growth parameter, and, ${\it
b}t $ - linearly time dependent cancer cells population decrease
"parameter". It can be considered that growth parameter ${\it a}$
is characteristic for type of the cancer cells and type of the
human individuals organs affected by cancer cells. Also, it can be
considered that decrease "parameter" ${\it b}t $  refers on the
decrease of the cancer cells attack efficiency (caused by mutation
by cancer cells) only.

As it is not hard to see cancer cells population (2) grows up till
time moment
\begin {equation}
  T = \frac {2{\it a}}{{\it b}}
\end {equation}

after which given population decreases toward zero (tends toward
zero in limit when t tends toward infinity). It means that for $t$
significantly larger than $T$ cancer cells realize a complete
self-destruction.

Also, according to (2), for two time moments $t_{1}$ and $t_{2}$
that satisfy condition $0 < t_{1}< t_{2}< T$, it follows
\begin {equation}
  p(t_{1}) < p(t_{2})                .
\end {equation}
It implies that quick (realized in a time interval much smaller
than $T$) reduction of the cancer cell population by some active
medical treatment (chirurgic treatment, radiation or
chemo-therapy) can be considered as a "time reduction" too. It,
formally speaking, forbids that time moment t become larger than
$T$ and, in this way, it forbids complete self-destruction of the
cancer cells. For this reason Brutovsky and Horvath strategy needs
counterintuitive complete rejection of any active medical
treatment.

A possible problem of the Brutovsky and Horvath strategy is that
cancer disease healing, except (2), needs a natural limitation
\begin {equation}
  p \leq p_{L}
\end {equation}
where $p_{L}$ represents critical cancer cells population that
almost certainly causes quick lethal effect. Also, there is
practically unambiguous empirical (clinical) fact that time moment
tL corresponding to $p_{L}$  is significantly larger than $T$,
i.e.
\begin {equation}
  t_{L} \ll T     .
\end {equation}
All this implies that in the real situations population dynamics
(1) must be reduced in the form
\begin {equation}
  \frac {dp}{dt} = {\it a}p     \hspace{1cm} {\rm for} \hspace{0.5 cm} t \leq t_{L}\ll T
\end {equation}
with simple solution
\begin {equation}
    p = p_{0}\exp[{\it a}t]  \hspace{1cm} {\rm for} \hspace{0.5 cm} t \leq t_{L}\ll T              .
\end {equation}
Obviously, in the limit when t tends toward $ t_{L}$, $p$ tends
toward $p_{L}$ corresponding to lethal finish.

So, it seems that in realistic situations we must use cancer cells
population dynamics (8).

A possible strategy against cancer disease with population
dynamics (8) holds, firstly, to discrete reduction of $p$ in a
smaller value $p_{R}$ by a fast (or discrete, i.e. with duration
significantly smaller than $ t_{L}$) and strong (with high
affectation at cancer cells and somewhat smaller affectation at
normal cells) active medical treatment. Given treatment is
realized by external (without human individual physiology domain),
physical or chemical means (chirurgic intervention, radiation or
injected chemo-therapy). This strategy holds additional continuous
(during whole time of the cancer cells population evolution)
reduction of ${\it a}$ in a smaller value ${\it a}_{R}$  by a not
so strong, active medical treatment. Given treatment (with not so
high affectation at cancer cells and somewhat smaller affectation
at normal cells) is realized by external (without human individual
physiology domain), physical or chemical means (oral
chemo-therapy) too.

Basic problem for mentioned strategy is, of course, its strong
character. Namely, for strong affectation at not only cancer but
also at the normal cells, strong, active treatment can be applied
only in a small time interval, i.e. quickly. In opposite case
given treatment itself can cause hard consequences including
patient dead. For this reason time interval necessary for reveal
of the normal cells before eventual repetition of strong treatment
can be too large in sense that it can admit a successful growth of
cancer cells by means of their population dynamics. On the other
hand, continuous, not so strong active treatment (oral chemo
therapy), for reason of its no so strong character (i.e. limited
efficiency), cannot reduce a in a smaller value ${\it a}_{R}$   in
sufficiently satisfactory way.

All this needs a new strategy with active treatment realized by
internal (within human individual physiology domain) means.
Namely, simply speaking, cancer originates when few cancer cells,
by developing a mimicry, becomes unrecognizable (as the defect
cells) for effector cells (citotoxic limfocits, nature killer
cells,  …). Then effector cells cannot destroy given cancer cells
which continue to reproduce unlimitedly. There are many attempts
of the stimulation of effector cells for a more sophisticated
recognition and elimination of the cancer cells as defect cells.
It, in fact, represents attempts for anti-cancer vaccine
development. But, it seems that anti-cancer vaccine must be
developed by any human individual, after moment when he obtains
cancer disease. It implies that today, by existing physical,
chemical and medical technology, such individual anti-cancer
vaccine can be developed during relatively large time interval
comparable with $t_{L}$ (There is opinion that individual
anti-cancer vaccine developing time can be much smaller in the
near future when nano-technology in medicine be better founded.)

All this implies that today there is no sufficiently satisfactory
strategy against cancer disease.

Nevertheless, we shall suggest an original strategy against cancer
disease that can be useful in at least some situations.

Consider a small time interval $\tau$ at end of which, i.e. in
time moment $\tau$, cancer cells population (8) can be linearly
approximated by
\begin {equation}
     p = p_{0} (1 + {\it a}\tau)
     \hspace{1cm} {\rm for} \hspace{0.5 cm}  {\it a}\tau \ll 1 \hspace{0.5 cm} {\rm and} \hspace{0.5 cm}  \tau \ll t_{L}          .
\end {equation}

Suppose that in the same moment a fast (for time interval
significantly smaller than $\tau$), internal active treatment
(whose nature will be discussed latter) can be realized.

Suppose that given treatment is half-strong. It means, on the one
hand, that given treatment reduce $p$ into $kp$ so that $k(1 +
{\it a}\tau)$ is smaller than 1, i.e.
\begin {equation}
     k(1 + {\it a}\tau) < 1   \hspace{1cm} {\rm for} \hspace{0.5 cm}{\it a}\tau \ll 1              .
\end {equation}
Or, by given treatment, cancer cells population (9) turns out
discretely in
\begin {equation}
     p_{R}= p_{0} k(1 + {\it a}\tau)  < p_{0}  \hspace{1cm} {\rm for} \hspace{0.5 cm}       {\it a}\tau \ll 1                 .
\end {equation}
On the other hand half-strong character of given treatment means
that after application of given treatment normal cells reveal
occurs during a time interval $\tau$.

All this implies that $n$ times repeated given fast, half-strong,
internal active treatment with period $\tau$ yields the following
cancer cells population
\begin {equation}
     p_{Rn}= p_{0}( k(1 + {\it a}\tau))^{n} < p_{0}  \hspace{1cm} {\rm for} \hspace{0.5 cm}       {\it a}\tau \ll 1
\end {equation}
where $n$ represents a natural number 1, 2, … . Obviously, (12)
tends toward zero when tends to be very large, or, formally
\begin {equation}
    Lim_{n \rightarrow \infty}p_{Rn} = 0   \hspace{1cm} {\rm for} \hspace{0.5 cm}      {\it a}\tau \ll 1     .
\end {equation}
But, really, $n$ can be relatively small if $k(1 + {\it a}\tau)$
is relatively small too. For example, for $k(1 + {\it a}\tau) =
0.8$ and $n=10$ it follows $p_{Rn}\simeq 0.1 p_{0}$ that
represents a very satisfactory result.

It is not hard to see that described fast, half-strong … treatment
conceptually corresponds to Zeno effect, theoretically predicted
[5] and experimentally verified [6], [7] in the quantum mechanics,
but existing in practically any domain of the physics [8].

Meanwhile, there is a principal question what really
physiologically can represent mentioned fast, half-strong …
treatment. We shall suggest a possible answer on this question.

In our previous work [4] it has been observed that cancer and
auto-immune diseases, e.g. multiple sclerosis, act, in some sense,
oppositely. It implies that a mechanism similar to auto immune
disease functioning can be used for fast, half-strong … treatment.

Concretely, suppose that system for effector cells activity
regulation,  precisely supression is inhibited by an external
chemical influence during a small time interval $\tau_{inh}$ many
times smaller than $\tau$. It will cause hyper-activity of the
effector cells (which will be called metaphorically dirty
inspector Harry effect) that internally, i.e. as the part of human
individual physiology, attack all cells, normal and cancer. There
is a probability that such attack can do larger damage of cancer
cells than normal cells.

Suppose further that after $\tau_{inh}$ inhibition of the
suppression cells is stopped, i.e. that after normal activity of
the suppression cells is restored during next time interval $\tau
$ .

It can be added that during $\tau $ suppression cells can be
additionally activated by external chemical simulation. It causes
additional suppression of the effector cells activity which,
maybe, can do a passiveness of the mimicry mechanism by cancer
cells. Then, by suppression cells inhibition and effector cells
hyper-activation during $\tau_{inh}$, elimination of the cancer
cells can be more efficient. In other words, we suppose that here,
maybe, the analogy with an especial therapy against chronic
hepatitis B exists. Namely, in some cases, before medicament
therapy against chronic hepatitis B, during a time interval there
is an external chemical suppression of all immune processes in the
human organism. After given suppression medicament therapy becomes
more efficient.

Finally, suppose that this alternation of the regulator cells
inhibition and normal activity is repeated many, i.e. n times,
where n represents a relatively large natural number.

In this way we obtain a concrete model of the fast, half-strong,
active internal treatment against cancer disease.

\section { Conclusion}

In conclusion we can shortly repeat and point out the following.
In this work we suggested an original strategy for cancer
inhibition and eventually elimination. It is based by frequent
(many times repeated) application of an especial active medical
treatment. This treatment represents such inhibition of the
regulator cells which cause super-activity of the effector cells
(citotoxic limfocits, nature killer cells,  …) that eliminate
cancer cells (metaphorically called dirty inspector Harry effect).
Conceptually, our strategy is similar to Zeno effect theoretically
predicted and experimentally verified in the quantum mechanics
(but which can be realized in practically any domain of the
physics).

\section { References}

\begin {itemize}

\item [[1]] B. Brutovsky, D. Horvath, {\it Optimization Aspects of Carcinogenesis}, q-bio.PE/0907.2004 and references therein
\item [[2]] N. L. Komarova, A. V. Sadovsky, F. Y. Wan, J. R. Social Interface {\bf 5} (2008) 105
\item [[3]] J. H. Holland, {\it Adaptation in Natural and Artificial Systems} (University of Michigan, Boston, 1975)
\item [[4]] V. Pankovi\'c, R. Glavatovi\'c, M. Krmar, {\it (Anti)Peter Principle - (Inverse) Discrete Logistic Equation with Imprecisely Estimated and Stimulated Carrying Capacity},        gen-ph/0907.5320
\item [[5]] B. Misra, C. J. G. Sudarshan, J. Math. Phys. {\bf 18} (1977) 756
\item [[6]] W. M. Itano, D. J. Heinsen, J. J. Bokkinger, D. J. Wineland, Phys. Rev. A {\bf 41} (1999) 2995
\item [[7]] M.C. Fischer, B.Gutierrez-Medina, M. G. Raizen, Phys. Rev. Lett. {\bf 87} (2001)  040402
\item [[8]] T. H$\ddot{u}$bsch, V. Pankovi\'c, {\it A Classical Switched LC/RC Circuit Modeling of the Quantum Zeno and anti-Zeno Effects},  quant-ph/0907.4361

\end{itemize}

\end{document}